\documentclass[12pt,english]{article}
\usepackage[T1]{fontenc}
\usepackage[latin9]{inputenc}
\usepackage{geometry}
\usepackage{color}
\usepackage{babel}
\usepackage{array}
\usepackage{booktabs}
\usepackage{url}
\usepackage{amstext}
\usepackage{graphicx}
\usepackage{setspace}
\usepackage[unicode=true,pdfusetitle,
 bookmarks=true,bookmarksnumbered=false,bookmarksopen=false,
 breaklinks=false,pdfborder={0 0 1},backref=false,colorlinks=true]
 {hyperref}
\hypersetup{
 urlcolor=blue}

\makeatletter

\providecommand{\tabularnewline}{\\}

\@ifundefined{date}{}{\date{}}
\makeatother

\begin{document}
\title{\textbf{Estimating the lifetime risk of a false positive screening
test result}}
\author{Tim White$^{*}$ and Sara Algeri\thanks{School of Statistics, University of Minnesota, Minneapolis, MN\protect \\
\protect \\
\hspace*{0.23in}\textit{Correspondence}: whit2618@umn.edu}}
\maketitle
\begin{abstract}
\hspace*{-0.225in}False positive results in screening tests have
potentially severe psychological, medical, and financial consequences
for the recipient. However, there have been few efforts to quantify
how the risk of a false positive accumulates over time. We seek to
fill this gap by estimating the probability that an individual who
adheres to the U.S. Preventive Services Task Force (USPSTF) screening
guidelines will receive at least one false positive in a lifetime.
To do so, we assembled a data set of 116 studies cited by the USPSTF
that report the number of true positives, false negatives, true negatives,
and false positives for the primary screening procedure for one of
five cancers or six sexually transmitted diseases. We use these data
to estimate the probability that an individual in one of 14 demographic
subpopulations will receive at least one false positive for one of
these eleven diseases in a lifetime. We specify a suitable statistical
model to account for the hierarchical structure of the data, and we
use the parametric bootstrap to quantify the uncertainty surrounding
our estimates. The estimated probability of receiving at least one
false positive in a lifetime is 85.5\% ($\pm$0.9\%) and 38.9\% ($\pm$3.6\%)
for baseline groups of women and men, respectively. It is higher for
subpopulations recommended to screen more frequently than the baseline,
including more vulnerable groups such as pregnant women and men who
have sex with men. Since screening technology is imperfect, false
positives remain inevitable. The high lifetime risk of a false positive
reveals the importance of educating patients about this phenomenon.\\
\end{abstract}
\begin{center}
\textit{\footnotesize{}Keywords}{\footnotesize{}: false positive,
screening test, lifetime risk, U.S. Preventive Services Task Force}{\footnotesize\par}
\par\end{center}

\vfill{}

\begin{singlespace}
\noindent \pagebreak{}
\end{singlespace}
\begin{singlespace}

\section{Introduction}
\end{singlespace}

\begin{doublespace}
Healthcare professionals encourage individuals to get screened regularly
for certain cancers, sexually transmitted diseases (STDs), and other
medical conditions. Programs of repeated screening are beneficial
because they allow for the early detection of these diseases, which,
in turn, increases the likelihood of successful treatment. However,
since even the most sophisticated modern screening technology falls
short of 100\% accuracy, some test results will inevitably be incorrect.
Specifically, the result of any screening test might wind up being
a false negative, indicating that the patient does not have the disease
when they do, or a false positive, indicating that the patient has
the disease when they do not.\\

Incorrect screening test results occur for a variety of reasons, ranging
from improper specimen collection to transcription and reporting inaccuracies.\textsuperscript{1}
Regardless of their cause, screening test errors have serious implications.
False negatives may delay the detection of potentially life-threatening
diseases, reduce public confidence in screening, and, in some cases,
induce legal action by the affected party.\textsuperscript{2} False
positives, on the other hand, take a toll on the recipient's mental
health, as they can generate stress and strain personal relationships.\textsuperscript{3-7}
There is also evidence that false positives reduce compliance with
subsequent screenings.\textsuperscript{8-9} Further, false positives
often require follow-up tests, and they can even prompt individuals
to undergo unnecessary and costly invasive procedures.\textsuperscript{10}\\

Despite these adverse psychological, medical, and financial effects,
efforts to quantify and communicate the risk of a false positive have
remained scarce. For most recommended procedures, reliable data exist
regarding the false positive rate of a single screening occasion.
However, these rates are far from common knowledge among the general
public. There have also been relatively few attempts to estimate the
probability of receiving at least one false positive when a particular
procedure is repeated over time.\textsuperscript{11-14} Even less
pursued is the estimation of this probability across multiple screening
procedures for different diseases.\textsuperscript{15-16} Still absent
in the literature is an effort to compute the probability of receiving
at least one false positive in a lifetime when an individual adheres
to a program of repeated screening for several diseases.\\

In this manuscript, we estimate the lifetime probability of a false
positive for individuals adhering to the screening guidelines of the
United States Preventive Services Task Force (USPSTF),\textsuperscript{17}
which are widely considered the gold standard among medical practitioners.
We consolidate the USPSTF recommendations and evidence into a data
set that reports the results of 116 studies, each of which summarizes
the accuracy of a screening procedure for one of five cancers or six
STDs. We use these data to estimate the probability of receiving at
least one false positive in a lifetime from any of the screening procedures
considered, and we replicate our analysis for 14 demographic and behavioral
subpopulations. Our findings provide patients and healthcare providers
with an individualized layer of information that sheds new light on
the long-term significance of screening test imprecision.\\

\end{doublespace}
\begin{onehalfspace}

\section{Methods}
\end{onehalfspace}
\begin{singlespace}

\subsection{Diseases and screening procedures}
\end{singlespace}

\begin{doublespace}
Many different procedures are used to screen individuals for cancers,
STDs, and other diseases like diabetes and osteoporosis. However,
most procedures are recommended only for individuals at high risk
for the disease, often only when risk factors emerge or intensify.
Our analysis focuses on the set of diseases for which routine screening
is advocated for large segments of the population. In addition, we
consider only those diseases for which the negative consequences of
false positives are well-documented or can be reasonably inferred
based on the existing literature.\\

We formalize these principles by using the following criteria to identify
the diseases relevant to our analysis: (1) The disease must be a cancer\textsuperscript{18}
or an STD\textsuperscript{19} and (2) the USPSTF must have assigned
a grade of C or higher to the screening service for the disease in
its most recent recommendation --- i.e., at minimum, it must recommend
the service to be offered selectively to patients based on their unique
circumstances. As detailed in Table S1 of the supplementary material,
there are five cancers (breast, cervical, colorectal, lung, and prostate)\textsuperscript{20-24}
and six STDs (chlamydia, gonorrhea, hepatitis B, hepatitis C, HIV,
and syphilis)\textsuperscript{25-31} that satisfy these criteria.\\

Some of these eleven diseases have only one recommended screening
modality, while others have several viable procedures. However, for
nearly all diseases where multiple screening procedures are available,
one is more common than the others in practice. With this in mind,
we focus on only the primary screening procedure for each disease.
The only two diseases for which the most common procedure is unclear
are HIV and syphilis; for these, we select the primary screening modality
based on the quantity and recency of available data.\\

\end{doublespace}
\begin{singlespace}

\subsection{Subpopulations and screening intervals}
\end{singlespace}

\begin{doublespace}
Not every individual is recommended to get screened for all eleven
diseases listed above. For each cancer and STD, the USPSTF guidelines
target only those segments of the population for which there are proven
benefits to screening. These groups are defined according to demographic
characteristics like age and sex or behavioral considerations like
sexual activity and smoking status.\\

We account for this heterogeneity in screening protocols by replicating
our analysis for 14 demographic and behavioral subpopulations. The
six female subpopulations are defined according to the individual's
smoking status and the number of pregnancies they expect to experience
in their lifetime. The eight male subpopulations are determined by
the individual's smoking status, whether or not they are a man who
has sex with men (MSM), and whether or not they intend to get screened
for prostate cancer, which is optional according to the USPSTF.\textsuperscript{24}
Baseline females are non-smokers who anticipate zero pregnancies in
their lifetime, while baseline males are non-smoking, non-MSM individuals
who do not intend to receive routine prostate exams. For both females
and males, we define smoking status according to the USPSTF eligibility
criteria for lung cancer screening --- i.e., smokers are individuals
with a 20 pack-year history who currently smoke or have quit within
the past 15 years.\textsuperscript{23}\\

Since the necessity and frequency of screening vary by subpopulation
for each disease, so too does the number of times that individuals
are recommended to get screened in a lifetime. For some diseases,
the lifetime number of screening occasions follows clearly from the
USPSTF guidelines. For others, the USPSTF guidelines lack either an
age range, an interval at which screening should be repeated, or both,
which makes it more complicated to derive the lifetime number of screening
occasions for each subpopulation. Section S1 explains our approach
to these more ambiguous cases, and Table S2 reports the assumed lifetime
number of screening occasions for each disease by subpopulation. Figure
1 lists the primary screening procedure and summarizes the USPSTF
guidelines for each disease considered in our analysis.\\

\end{doublespace}
\begin{singlespace}

\subsection{Data collection}
\end{singlespace}

\begin{doublespace}
Given the number of times an individual is recommended to get screened
for a particular disease in a lifetime and provided that we have an
estimate of the false positive rate for each screening occasion, we
can estimate the probability that at least one of these occasions
will result in a false positive. Let $p_{d}$ denote the probability
that an individual tests positive for disease $d$ when the disease
is not present. Our first objective is to obtain an estimate, $\widehat{p}_{d}$,
of $p_{d}$.\\

\begin{figure}
\textbf{\caption{Timeline of screening guidelines}
}

\medskip{}

\noindent \hspace*{-0.1in}\includegraphics{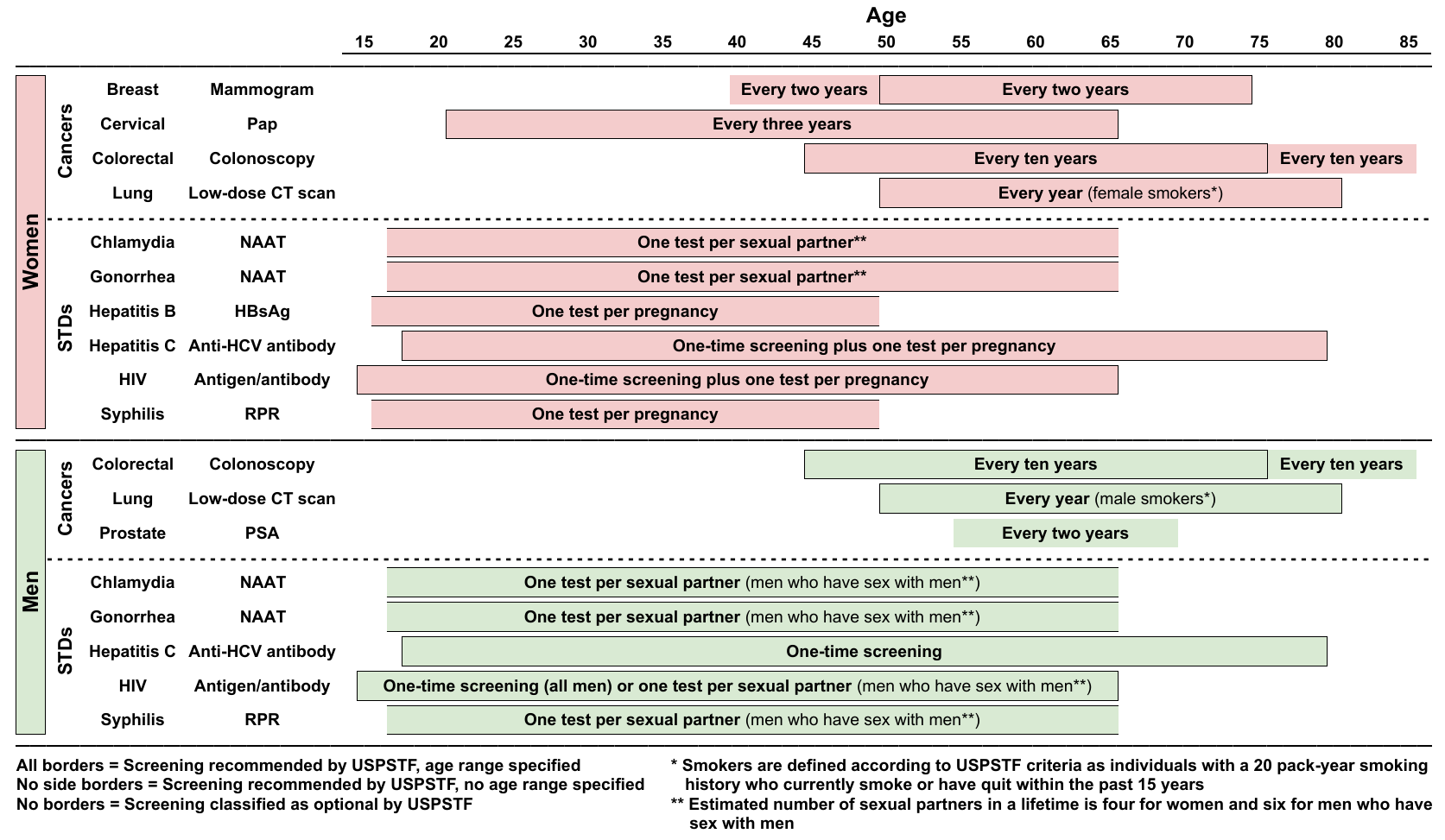}
\end{figure}

To compile the data necessary to obtain $\widehat{p}_{d}$, we first
identified the most recent USPSTF recommendation statement for each
disease as of August 31, 2021 (see Table S1). Next, we extracted all
sources among those cited by the USPSTF that reported the number of
true positives, false negatives, true negatives, and false positives
from a study involving the primary screening procedure for that disease.
If no such studies were cited in the USPSTF recommendation statement
for a particular disease, we referred to the corresponding USPSTF
evidence review and proceeded similarly. In cases where neither of
the two current documents contained relevant studies, we applied the
same strategy to the next most recent set of USPSTF guidelines. See
Table S3 for more details about the data collection procedure for
each disease.\\

This process yielded 116 studies. Let $S_{d}$ denote the number of
studies collected for disease $d$ using the above methodology. The
estimated false positive rate of the primary screening procedure for
disease $d$ is given by

\begin{equation}
\widehat{p}_{d}=\frac{\sum_{s=1}^{S_{d}}\text{FP}_{s}}{\sum_{s=1}^{S_{d}}n_{s}},
\end{equation}

where $\text{FP}_{s}$ is the number of false positives in study $s\in\{1,...,S_{d}\}$
and $n_{s}$ is the number of false positives plus the number of true
negatives. It follows from (1) that $\widehat{p}_{d}$ can be used
to estimate the conditional probability of testing positive for disease
$d$ given that the disease is not present in the individual ---
i.e., the false positive probability of a single screening occasion
for disease $d$. Table 1 reports the estimates obtained from (1)
for each disease. Table 2 summarizes the notation used in Sections
2.3 and 2.4.\\

\end{doublespace}
\begin{singlespace}

\subsection{Statistical methods}
\end{singlespace}

\begin{doublespace}
We seek to model the probability $p_{i}$ that a healthy individual
in subpopulation $i$ will receive at least one false positive in
a lifetime for at least one of the five cancers or six STDs considered,
assuming that they adhere to the USPSTF screening guidelines. We define
healthy individuals as those who remain negative for all five cancers
and six STDs throughout their lifetime.\\

\begin{table}[t]
\textbf{\caption{Estimated false positive probability of one screening occasion}
}

\medskip{}

\begin{singlespace}
\centering{}{\footnotesize{}}%
\begin{tabular}{ccc}
\toprule 
\textbf{\footnotesize{}Disease} & \textbf{\footnotesize{}Screening procedure} & \textbf{\footnotesize{}Estimate (SE)}\tabularnewline
\midrule 
{\footnotesize{}Breast cancer} & {\footnotesize{}Mammogram} & {\footnotesize{}4.9\% (0.1\%)}\tabularnewline
\midrule 
{\footnotesize{}Cervical cancer} & {\footnotesize{}Pap test} & {\footnotesize{}5.0\% (0.1\%)}\tabularnewline
\midrule 
{\footnotesize{}Chlamydia} & {\footnotesize{}NAAT} & {\footnotesize{}0.5\% (<0.1\%)}\tabularnewline
\midrule 
{\footnotesize{}Colorectal cancer} & {\footnotesize{}Colonoscopy} & {\footnotesize{}11.3\% (1.3\%)}\tabularnewline
\midrule 
{\footnotesize{}Gonorrhea} & {\footnotesize{}NAAT} & {\footnotesize{}0.2\% (<0.1\%)}\tabularnewline
\midrule 
{\footnotesize{}Hepatitis B} & {\footnotesize{}HBsAg test} & {\footnotesize{}2.0\% (0.1\%)}\tabularnewline
\midrule 
{\footnotesize{}Hepatitis C} & {\footnotesize{}Anti-HCV antibody test} & {\footnotesize{}1.0\% (0.2\%)}\tabularnewline
\midrule 
{\footnotesize{}HIV} & {\footnotesize{}Antigen/antibody test} & {\footnotesize{}0.2\% (<0.1\%)}\tabularnewline
\midrule 
{\footnotesize{}Lung cancer} & {\footnotesize{}Low-dose CT scan} & {\footnotesize{}20.7\% (0.1\%)}\tabularnewline
\midrule 
{\footnotesize{}Prostate cancer} & {\footnotesize{}PSA test} & {\footnotesize{}10.2\% (0.3\%)}\tabularnewline
\midrule 
{\footnotesize{}Syphilis} & {\footnotesize{}RPR test} & {\footnotesize{}0.3\% (<0.1\%)}\tabularnewline
\bottomrule
\end{tabular}{\footnotesize\par}
\end{singlespace}
\end{table}

In order to model $p_{i}$, we first model the probability $P_{id}$
that a healthy individual in subpopulation $i$ will receive at least
one false positive for disease $d$ in a lifetime, assuming that they
get screened the recommended number of times $T_{id}$ with the primary
screening procedure for disease $d$. We also assume that, for each
disease $d$, the results of each screening occasion are independent
from the results of the previous occasions. Under these assumptions,
the probability $P_{id}$ can be expressed as

\vspace*{-0.25in}
\begin{equation}
P_{id}=1-(1-p_{d})^{T_{id}}.
\end{equation}

\end{doublespace}

\smallskip{}

\begin{doublespace}
Next, we extend this formula to incorporate more than one disease.
Let $\mathcal{D}_{i}$ denote the set of diseases for which an individual
in subpopulation $i$ is recommended to get screened at least once.
We assume that the event of receiving at least one false positive
in a lifetime for each disease $d\in\mathcal{D}_{i}$ is independent
from the same event for each of the other diseases in $\mathcal{D}_{i}$.
It follows from this assumption that the probability $p_{i}$, as
defined above, is given by

\vspace*{-0.2in}
\begin{equation}
p_{i}=1-\prod_{d\in\mathcal{D}_{i}}(1-P_{id})=1-\prod_{d\in\mathcal{D}_{i}}(1-p_{d})^{T_{id}}.
\end{equation}

\end{doublespace}

\bigskip{}

\begin{doublespace}
The derivations of (2) and (3) are provided in Sections S2 and S3,
respectively. These equations are informative in that they describe
the relationship between $p_{d}$, $P_{id}$, and $p_{i}$, but they
are limited by the fact that these quantities are unknown in the real
world. However, we can estimate (2) and (3) by plugging in the estimates
$\widehat{p}_{d}$ from (1). We obtain

\vspace*{-0.2in}
\begin{equation}
\widehat{P}_{id}=1-(1-\widehat{p}_{d})^{T_{id}}\qquad\text{and}\qquad\widehat{p}_{i}=1-\prod_{d\in\mathcal{D}_{i}}(1-\widehat{P}_{id})=1-\prod_{d\in\mathcal{D}_{i}}(1-\widehat{p}_{d})^{T_{id}},
\end{equation}

where the latter can be used to estimate, as desired, the probability
that a healthy individual in subpopulation $i$ will receive at least
one false positive in a lifetime.\\

\begin{table}
\textbf{\caption{Notation and definitions}
}

\medskip{}

\noindent \centering{}\hspace*{-0.225in}%
\begin{tabular}{>{\raggedleft}p{0.25in}>{\raggedright}p{3in}>{\raggedleft}p{0.25in}>{\raggedright}p{2.8in}}
\toprule 
{\scriptsize{}$S_{d}$} & {\scriptsize{}Number of studies collected for disease $d$} & {\scriptsize{}$n_{s}$} & {\scriptsize{}$TN_{s}+FP_{s}$}\tabularnewline
{\scriptsize{}$TP_{s}$} & {\scriptsize{}Number of true positives observed in study $s$} & {\scriptsize{}$N_{s}$} & {\scriptsize{}$TP_{s}+FN_{s}+TN_{s}+FP_{s}$}\tabularnewline
{\scriptsize{}$FN_{s}$} & {\scriptsize{}Number of false negatives observed in study $s$} & {\scriptsize{}$\widehat{p}_{FP,s}$} & {\scriptsize{}Proportion of cases in study $s$ that are false positives}\tabularnewline
{\scriptsize{}$TN_{s}$} & {\scriptsize{}Number of true negatives observed in study $s$} & {\scriptsize{}$\widehat{p}_{TN,s}$} & {\scriptsize{}Proportion of cases in study $s$ that are true negatives}\tabularnewline
{\scriptsize{}$FP_{s}$} & {\scriptsize{}Number of false positives observed in study $s$} & {\scriptsize{}$\widehat{p}_{+,s}$} & {\scriptsize{}Proportion of cases in study $s$ that are true positives
or false negatives}\tabularnewline
 & \multicolumn{3}{>{\raggedright}p{6.25in}}{}\tabularnewline
{\scriptsize{}$Z_{s}$} & \multicolumn{3}{>{\raggedright}p{6.25in}}{{\scriptsize{}Multinomial random variable representing the number
of true positives, false negatives, true negatives, and false positives
in study $s$}}\tabularnewline
{\scriptsize{}$B$} & \multicolumn{3}{>{\raggedright}p{6.25in}}{{\scriptsize{}Number of bootstrap iterations}}\tabularnewline
 & \multicolumn{3}{l}{}\tabularnewline
{\scriptsize{}$p_{d}$} & \multicolumn{3}{>{\raggedright}p{6.25in}}{{\scriptsize{}Probability of receiving a false positive on one screening
occasion for disease $d$}}\tabularnewline
{\scriptsize{}$T_{id}$} & \multicolumn{3}{>{\raggedright}p{6.25in}}{{\scriptsize{}Number of times an individual in subpopulation $i$
is recommended to get screened for disease $d$ in a lifetime}}\tabularnewline
{\scriptsize{}$P_{id}$} & \multicolumn{3}{>{\raggedright}p{6.25in}}{{\scriptsize{}Probability that a healthy individual in subpopulation
$i$ who adheres to the USPSTF screening guidelines will receive at
least one false positive for disease $d$ in a lifetime}}\tabularnewline
{\scriptsize{}$\mathcal{D}_{i}$} & \multicolumn{3}{>{\raggedright}p{6.25in}}{{\scriptsize{}Set of diseases for which an individual in subpopulation
$i$ is recommended to get screened at least once in a lifetime}}\tabularnewline
{\scriptsize{}$p_{i}$} & \multicolumn{3}{>{\raggedright}p{6.25in}}{{\scriptsize{}Probability that a healthy individual in subpopulation
$i$ who adheres to the USPSTF screening guidelines will receive at
least one false positive for any of the diseases in $\mathcal{D}_{i}$
in a lifetime}}\tabularnewline
\bottomrule
\end{tabular}
\end{table}

Finally, to quantify the uncertainty surrounding our estimators, we
rely on the assumption that the number of true positives (TP), false
negatives (FN), true negatives (TN), and false positives (FP) from
each study $s$ can be modeled by a multinomial random variable $Z_{s}\sim\text{Multinomial}(N_{s}\text{ },\text{ }\widehat{p}_{FP,s}\text{ },\text{ }\widehat{p}_{TN,s}\text{ },\text{ }\widehat{p}_{+,s})$,
where $N_{s}$ is the total sample size of study $s$, $\widehat{p}_{FP,s}$
is the proportion of cases in study $s$ that are false positives,
$\widehat{p}_{TN,s}$ is the proportion that are true negatives, and
$\widehat{p}_{+,s}$ is the proportion that are true positives or
false negatives. We incorporate the multinomial errors into our analysis
by means of the parametric bootstrap.\textsuperscript{32} Specifically,
for each subpopulation $i$, we simulate from the $\text{Multinomial}(N_{s}\text{ },\text{ }\widehat{p}_{FP,s}\text{ },\text{ }\widehat{p}_{TN,s}\text{ },\text{ }\widehat{p}_{+,s})$
distribution to account for the inherent randomness of the results
of each study $s\in\{1,...,S_{d}\}$ for each disease $d\in\mathcal{D}_{i}$.
We plug these simulated results into the equations for $\widehat{p}_{d}$
and $\widehat{P}_{id}$ to obtain one realization of $\widehat{p}_{d}$
and $\widehat{P}_{id}$ for each disease $d\in\mathcal{D}_{i}$, as
well as one realization of $\widehat{p}_{i}$. We carry out this process
$B=10,000$ times to obtain 10,000 realizations of each estimator,
and we use these realizations to compute their respective standard
errors. For $\widehat{p}_{i}$, the standard error is given by

\vspace*{-0.1in}
\begin{equation}
SE(\widehat{p}_{i})=\sqrt{\frac{1}{B-1}\sum_{b=1}^{B}(\widehat{p}_{i}^{\text{ }(b)}-\overline{\widehat{p}}_{i})^{2}},
\end{equation}

where $\overline{\widehat{p}}_{i}=\frac{1}{B}\sum_{b=1}^{B}\widehat{p}_{i}^{\text{ }(b)}$.
We use an analogous formula for $SE(\widehat{p}_{d})$ and $SE(\widehat{P}_{id})$.
These standard errors are reported along with the corresponding point
estimates.\\

\end{doublespace}
\begin{singlespace}

\section{Results}
\end{singlespace}

\begin{doublespace}
For each subpopulation $i$, Table 3 reports the estimated probability
$\widehat{p}_{i}$ that a healthy individual adhering to the USPSTF
screening guidelines will receive at least one false positive in a
lifetime for one of the five cancers or six STDs considered. Table
S4 reports the lifetime probability of a false positive separately
for the cancers and the STDs, and Table S5 breaks these probabilities
down by disease for each subpopulation by reporting the estimates
$\widehat{P}_{id}$ from (4).\\

\begin{table}[t]
\textbf{\caption{Estimated lifetime false positive probability by subpopulation}
}

\medskip{}

\begin{singlespace}
\centering{}{\footnotesize{}}%
\begin{tabular}{cc}
\toprule 
\textbf{\footnotesize{}Subpopulation} & \textbf{\footnotesize{}Estimate (SE)}\tabularnewline
\midrule 
{\footnotesize{}Baseline females} & {\footnotesize{}85.5\% (0.9\%)}\tabularnewline
\midrule 
{\footnotesize{}Females, one pregnancy} & {\footnotesize{}86.0\% (0.8\%)}\tabularnewline
\midrule 
{\footnotesize{}Females, two pregnancies} & {\footnotesize{}86.5\% (0.8\%)}\tabularnewline
\midrule 
{\footnotesize{}Female smokers} & {\footnotesize{}88.5\% (0.7\%)}\tabularnewline
\midrule 
{\footnotesize{}Female smokers, one pregnancy} & {\footnotesize{}88.9\% (0.7\%)}\tabularnewline
\midrule 
{\footnotesize{}Female smokers, two pregnancies} & {\footnotesize{}89.3\% (0.6\%)}\tabularnewline
\midrule 
{\footnotesize{}Baseline males} & {\footnotesize{}38.9\% (3.6\%)}\tabularnewline
\midrule 
{\footnotesize{}Men who have sex with men (MSM)} & {\footnotesize{}43.1\% (3.4\%)}\tabularnewline
\midrule 
{\footnotesize{}Male smokers} & {\footnotesize{}51.5\% (2.9\%)}\tabularnewline
\midrule 
{\footnotesize{}MSM smokers} & {\footnotesize{}54.9\% (2.7\%)}\tabularnewline
\midrule 
{\footnotesize{}Males, routine prostate exams} & {\footnotesize{}74.2\% (1.7\%)}\tabularnewline
\midrule 
{\footnotesize{}MSM, routine prostate exams} & {\footnotesize{}76.0\% (1.6\%)}\tabularnewline
\midrule 
{\footnotesize{}Male smokers, routine prostate exams} & {\footnotesize{}79.6\% (1.3\%)}\tabularnewline
\midrule 
{\footnotesize{}MSM smokers, routine prostate exams} & {\footnotesize{}81.0\% (1.2\%)}\tabularnewline
\bottomrule
\end{tabular}{\footnotesize\par}
\end{singlespace}
\end{table}

The estimated lifetime probability of a false positive is at least
38\% for all subpopulations and is greater than 50\% for all but two.
It exceeds 85\% for all female subpopulations and is only slightly
higher for female smokers (88.5\% $\pm$ 0.7\%) and pregnant women
(86.0\% $\pm$ 0.8\%) than baseline females (85.5\% $\pm$ 0.9\%).
This similarity in lifetime false positive risk between baseline and
non-baseline females stems from the fact that women have a high probability
of receiving at least one false positive in a lifetime for breast
cancer and cervical cancer relative to the other cancers and STDs
(see Table S5), and the USPSTF guidelines for these two diseases are
the same across all female subpopulations. The USPSTF guidelines for
breast cancer and cervical cancer prescribe more screening occasions
in a lifetime than the guidelines for the other cancers and STDs (see
Table S2), so it is not surprising that these two diseases contribute
so prominently to lifetime false positive risk among females.\\

There is far more variability in lifetime false positive risk among
males than females, as the estimated lifetime probability of a false
positive ranges from around 39\% for baseline males (38.9\% $\pm$
3.6\%) to more than 80\% for MSM smokers who elect to get screened
for prostate cancer (81.0\% $\pm$ 1.2\%). This variation is attributable
to the fact that the USPSTF screening guidelines differ substantially
across the male subpopulations. For instance, while baseline males
are only recommended to get screened for one cancer and two STDs,
MSM smokers who receive routine prostate exams are recommended to
get screened for three cancers and five STDs (see Table S2).\\

We can use odds ratios to compare the lifetime risk of a false positive
between subpopulations. The odds ratio for two subpopulations $i_{1}$
and $i_{2}$ is given by $(\widehat{p}_{i_{1}}/(1-\widehat{p}_{i_{1}}))/(\widehat{p}_{i_{2}}/(1-\widehat{p}_{i_{2}}))$,
where $\widehat{p}_{i_{1}}$ and $\widehat{p}_{i_{2}}$ are the respective
lifetime false positive probabilities for the two subpopulations.
Applying this formula to the estimates in Table 3, we find that the
odds of receiving at least one false positive in a lifetime are 9.30
times higher for baseline females than baseline males and 7.27 times
higher for female smokers than male smokers. Much of the discrepancy
in lifetime false positive risk between females and males can be explained
by the high lifetime false positive probability for breast cancer
and cervical cancer, as women are recommended to get screened for
these two diseases while men are not.\\

Among females, the odds of receiving at least one false positive in
a lifetime are 1.30 times higher for female smokers than baseline
females. Relative to baseline females, the odds are 1.04 times higher
for women who experience one pregnancy and 1.08 times higher for women
who experience two pregnancies. These odds ratios are consistent with
the small amount of variation in lifetime false positive risk among
the female subpopulations.\\

Relative to baseline males, the odds of receiving at least one false
positive in a lifetime are 1.19 times higher for men who have sex
with men, 1.67 times higher for male smokers, and 4.53 times higher
for males who undergo routine prostate exams. This last result suggests
that the decision to get screened routinely for prostate cancer has
a considerable effect on lifetime false positive risk among males,
which is not surprising considering the high estimated probability
of receiving a false positive from one prostate-specific antigen (PSA)
test (see Table 1) and the relatively frequent time interval for prostate
cancer screening (see Figure 1 and Table S2).\\

\end{doublespace}
\begin{singlespace}

\section{Discussion}
\end{singlespace}

\begin{doublespace}
To our knowledge, this manuscript is the first to quantify the lifetime
risk of a false positive for individuals adhering to a program of
repeated screening for multiple diseases. Our results indicate that
healthy individuals who follow the USPSTF screening guidelines for
a particular set of five cancers and six STDs have a high probability
of receiving at least one false positive in a lifetime. Notably, the
estimated lifetime probability of a false positive is high relative
to the baseline for more vulnerable groups such as pregnant women
and men who have sex with men. Efforts by healthcare providers to
communicate the lifetime risk of a false positive are likely to be
particularly beneficial for these groups.\\

Our methodology relies on several assumptions about screening procedures,
participants, and intervals. While our conclusions are robust to minor
violations of these assumptions, we recognize that some of them may
not hold in practice for all individuals. First, the probabilities
estimated in this manuscript are conditional on the screening participant
remaining negative for all five cancers and six STDs throughout their
lifetime. The estimates reported in Tables 1, 3, S4, and S5 are not
valid for individuals who contract one or more of the eleven diseases
throughout their lifetime.\\

Second, the accuracy of our estimates depends on the extent to which
patients and healthcare providers adhere to the USPSTF screening guidelines.
We maintain that our assumption of strict adherence to the USPSTF
guidelines is the most defensible way to determine the number of times
an individual is recommended to get screened for a particular disease
in a lifetime. Nevertheless, several studies suggest that the implementation
of these guidelines in clinical settings is imperfect.\textsuperscript{33-35}
As such, the values reported in Table S2 may overestimate or underestimate
the number of times individuals actually get screened for certain
diseases in a lifetime.\\

Third, our assumption that individuals receive only the primary screening
procedure for each disease likely does not capture the nuances of
actual screening practices, which can vary across healthcare systems.
Multiple screening procedures are available for several of the diseases
considered in our analysis, and while one procedure tends to be more
common than the others for each disease, the proportion of individuals
that get screened with the other modalities may be nontrivial.\\

Finally, our determination of the lifetime number of screening occasions
for the STDs relies on a few assumptions that may oversimplify the
individualized nature of STD screening. The USPSTF guidelines for
STD screening are highly contingent on risk factors and personal circumstances,
and they generally do not provide both an age range and a time interval
at which screening procedures should be repeated.\textsuperscript{25-31}
It is therefore possible, if not likely, that two individuals in the
same subpopulation who both adhere to the USPSTF guidelines will get
screened a different number of times for a particular STD in a lifetime.
While it would be reasonable to estimate the lifetime number of STD
screening occasions for each subpopulation empirically, the data required
for this task are not readily available.\\

Even with these limitations in mind, our results offer undeniable
evidence that the long-term burden of screening test imprecision is
worth the attention of patients and healthcare providers. Specifically,
our findings reveal the importance of communicating the pervasiveness
of false positive results to patients. This transparency is intended
not to sow distrust in screening procedures or the guidelines associated
with them, but rather to improve the ability of patients to respond
sensibly to an unexpected positive result. An individual who is informed
about the estimated lifetime false positive probability for someone
with their demographic and behavioral characteristics is likely to
exhibit a more measured response in this scenario than an uninformed
individual.\\

To aid in this endeavor, we disseminate our findings via a web application
called The False Positives Calculator.\textsuperscript{36} This interactive
tool, which is openly accessible at \url{https://falsepositives.shinyapps.io/calculator},
allows users to extract personalized information about screening test
accuracy and learn about the USPSTF guidelines for the eleven diseases
considered in our analysis. It also hosts our data set (free for download
at \url{https://github.com/timwhite0/false-positives-calculator}),
which, to the best of our knowledge, is the largest centralized, publicly
available data source on screening test accuracy.\\

False positive results are an inevitable feature of screening tests,
as they cannot be eliminated without simultaneously increasing the
prevalence of false negatives. For as long as screening technology
remains imperfect, the best course of action for patients and healthcare
providers is to remain informed about the risk of a false positive,
paying particular attention to how this risk accumulates over time.
This manuscript provides a framework for quantifying the lifetime
risk of a false positive, one that could feasibly be updated as screening
technology evolves and more data become available.
\end{doublespace}

\begin{singlespace}
\noindent \pagebreak{}
\end{singlespace}


\begin{thebibliography}{10}
\begin{singlespace}
\bibitem{key-18}Wians FH Jr. Clinical laboratory tests: Which, why,
and what do the results mean? Lab Med. 2009; 40: 105-113.

\bibitem{key-72}Petticrew MP, Sowden AJ, Lister-Sharp D, Wright K.
False-negative results in screening programmes: systematic review
of impact and implications. Health Technol Assess. 2000; 4: 1-120.

\bibitem{key-35}Fowler FJ Jr, Barry MJ, Walker-Corkery B, Caubet
J-F, Bates DW, Lee JM, et al. The impact of a suspicious prostate
biopsy on patients\textquoteright{} psychological, socio-behavioral,
and medical care outcomes. J Gen Intern Med. 2006; 21: 715-721.

\bibitem{key-20}Katz AR, Effler PV, Ohye RG, Brouillet B, Lee MVC,
Whiticar PM. False-positive gonorrhea test results with a nucleic
acid amplification test: the impact of low prevalence on positive
predictive value. Clin Infect Dis. 2004; 38: 814-819.

\bibitem{key-21}Salz T, Gottlieb SL, Smith JS, Brewer NT. The association
between cervical abnormalities and attitudes toward cervical cancer
prevention. J Womens Health (Larchmt). 2010; 19: 2011-2016.

\bibitem{key-22}Shanks L, Klarkowski D, O\textquoteright Brien DP.
False positive HIV diagnoses in resource limited settings: operational
lessons learned for HIV programmes. PLoS One. 2013; 8: e59906.

\bibitem{key-23}Toft EL, Kaae SE, Malmqvist J, Brodersen J. Psychosocial
consequences of receiving false-positive colorectal cancer screening
results: a qualitative study. Scand J Prim Health Care. 2019; 37:
145-154.

\bibitem{key-25}Bond M, Pavey T, Welch K, Cooper C, Garside R, Dean
S, et al. Systematic review of the psychological consequences of false-positive
screening mammograms. Health Technol Assess. 2013; 17: 1-170, v-vi.

\bibitem{key-26}Ford ME, Havstad SL, Flickinger L, Johnson CC. Examining
the effects of false positive lung cancer screening results on subsequent
lung cancer screening adherence. Cancer Epidemiol Biomarkers Prev.
2003; 12: 28-33.

\bibitem{key-27}Lidbrink E, Elfving J, Frisell J, Jonsson E. Neglected
aspects of false positive findings of mammography in breast cancer
screening: analysis of false positive cases from the Stockholm trial.
BMJ. 1996; 312: 273-276.

\bibitem{key-29}Hubbard RA, Kerlikowske K, Flowers CI, Yankaskas
BC, Zhu W, Miglioretti DL. Cumulative probability of false-positive
recall or biopsy recommendation after 10 years of screening mammography:
a cohort study. Ann Intern Med. 2011; 155: 481-492.

\bibitem{key-32}Njor SH, Olsen AH, Schwartz W, Vejborg I, Lynge E.
Predicting the risk of a false-positive test for women following a
mammography screening programme. J Med Screen. 2007; 14: 94-97.

\bibitem{key-30}Croswell JM, Baker SG, Marcus PM, Clapp JD, Kramer
BS. Cumulative incidence of false-positive test results in lung cancer
screening: a randomized trial. Ann Intern Med. 2010; 152: 505-512.

\bibitem{key-45}Elmore JG, Barton MB, Moceri VM, Polk S, Arena PJ,
Fletcher SW. Ten-year risk of false positive screening mammograms
and clinical breast examinations. N Engl J Med. 1998; 338: 1089-1096.

\bibitem{key-34}Baker SG, Kramer BS. Estimating the cumulative risk
of a false-positive under a regimen involving various types of cancer
screening tests. J Med Screen. 2008; 15: 18-22.

\bibitem{key-35}Croswell JM, Kramer BS, Kreimer AR, Prorok PC, Xu
J-L, Baker SG, et al. Cumulative incidence of false-positive results
in repeated, multimodal cancer screening. Ann Fam Med. 2009; 7: 212-222.

\bibitem{key-1}U.S. Preventive Services Task Force. Home Page. {[}cited
15 Jun 2022{]}. Available from: https://www.uspreventiveservicestaskforce.org/uspstf/

\bibitem{key-2}National Cancer Institute. Cancer Types. {[}cited
15 Jun 2022{]}. Available from: https://www.cancer.gov/types

\bibitem{key-70}Centers for Disease Control and Prevention. Diseases
\& Related Conditions. {[}cited 15 Jun 2022{]}. Available from: https://www.cdc.gov/std/general/default.htm

\bibitem{key-4}Siu AL, U.S. Preventive Services Task Force. Screening
for breast cancer: U.S. Preventive Services Task Force recommendation
statement. Ann Intern Med. 2016; 164: 279-296.

\bibitem{key-5}Curry SJ, Krist AH, Owens DK, Barry MJ, Caughey AB,
Davidson KW, et al. Screening for cervical cancer: US Preventive Services
Task Force recommendation statement. JAMA. 2018; 320: 674-686.

\bibitem{key-6}Davidson KW, Barry MJ, Mangione CM, Cabana M, Caughey
AB, Davis EM, et al. Screening for colorectal cancer: US Preventive
Services Task Force recommendation statement. JAMA. 2021; 325: 1965-1977.

\bibitem{key-7}Krist AH, Davidson KW, Mangione CM, Barry MJ, Cabana
M, Caughey AB, et al. Screening for lung cancer: US Preventive Services
Task Force recommendation statement. JAMA. 2021; 325: 962-970.

\bibitem{key-8}Grossman DC, Curry SJ, Owens DK, Bibbins-Domingo K,
Caughey AB, Davidson KW, et al. Screening for prostate cancer: US
Preventive Services Task Force recommendation statement. JAMA. 2018;
319: 1901-1913.

\bibitem{key-9}LeFevre ML, U.S. Preventive Services Task Force. Screening
for chlamydia and gonorrhea: U.S. Preventive Services Task Force recommendation
statement. Ann Intern Med. 2014; 161: 902-910.

\bibitem{key-10}Krist AH, Davidson KW, Mangione CM, Barry MJ, Cabana
M, Caughey AB, et al. Screening for hepatitis B virus infection in
adolescents and adults: US Preventive Services Task Force recommendation
statement. JAMA. 2020; 324: 2415-2422.

\bibitem{key-16}Owens DK, Davidson KW, Krist AH, Barry MJ, Cabana
M, Caughey AB, et al. Screening for hepatitis B virus infection in
pregnant women: US Preventive Services Task Force reaffirmation recommendation
statement. JAMA. 2019; 322: 349-354.

\bibitem{key-17}Owens DK, Davidson KW, Krist AH, Barry MJ, Cabana
M, Caughey AB, et al. Screening for hepatitis C virus infection in
adolescents and adults: US Preventive Services Task Force recommendation
statement. JAMA. 2020; 323: 970-975.

\bibitem{key-18}Owens DK, Davidson KW, Krist AH, Barry MJ, Cabana
M, Caughey AB, et al. Screening for HIV infection: US Preventive Services
Task Force recommendation statement. JAMA. 2019; 321: 2326-2336.

\bibitem{key-19}Bibbins-Domingo K, Grossman DC, Curry SJ, Davidson
KW, Epling JW Jr, García FA, et al. Screening for syphilis infection
in nonpregnant adults and adolescents: US Preventive Services Task
Force recommendation statement. JAMA. 2016; 315: 2321-2327.

\bibitem{key-20}Curry SJ, Krist AH, Owens DK, Barry MJ, Caughey AB,
Davidson KW, et al. Screening for syphilis infection in pregnant women:
US Preventive Services Task Force reaffirmation recommendation statement.
JAMA. 2018; 320: 911-917.

\bibitem{key-48}Efron B, Tibshirani RJ. An introduction to the bootstrap.
Chapman and Hall/CRC; 1994.

\bibitem{key-32}Corbelli J, Borrero S, Bonnema R, McNamara M, Kraemer
K, Rubio D, et al. Physician adherence to U.S. Preventive Services
Task Force mammography guidelines. Womens Health Issues. 2014; 24:
e313-e319.

\bibitem{key-33}Nelson W, Moser RP, Gaffey A, Waldron W. Adherence
to cervical cancer screening guidelines for U.S. women aged 25-64:
data from the 2005 Health Information National Trends Survey (HINTS).
J Womens Health (Larchmt). 2009; 18: 1759-1768.

\bibitem{key-34}Cyhaniuk A, Coombes ME. Longitudinal adherence to
colorectal cancer screening guidelines. Am J Manag Care. 2016; 22:
105-111.

\bibitem{key-50}White T. The False Positives Calculator. 2021 {[}cited
15 Jun 2022{]}. Available from: https://falsepositives.shinyapps.io/calculator
\end{singlespace}
\end{thebibliography}
\end{document}